\documentclass[12pt]{iopart}

\usepackage[figuresright]{rotating}
\usepackage{psfig}
\usepackage{epsfig}
\usepackage{floatflt}
\usepackage{graphicx}
\usepackage{dcolumn}
\usepackage{bm}

\newcommand{\AmS}{{\protect\the\textfont2
  A\kern-.1667em\lower.5ex\hbox{M}\kern-.125emS}}

\def\Journal#1#2#3#4{{#1} {\bf #2}, #3 (#4)}

\def\NPA{{ Nucl. Phys.} A}
\def\PLB{{ Phys. Lett.}  B}
\def\PRL{ Phys. Rev. Lett.}

\def\PRC{{ Phys. Rev.} C}
\def\PRD{{ Phys. Rev.} D}

\begin{document}

\title[Open Charm Production in $\sqrt{s_{NN}}$=200 GeV Au+Au
Collisions]{Open Charm Production in $\sqrt{s_{NN}}$=200 GeV Au+Au
Collisions}

\author{Yifei Zhang (for the STAR\footnote[1]{For the full list
of STAR authors and acknowledgements, see appendix `Collaboration'
of this volume.} Collaboration)}
\address{Dept. of Modern Physics, University of Science and Technology of China, Hefei, Anhui, China, 230026}
\address{Lawrence Berkeley National Laboratory, 1 Cyclotron Road, MS70R319, Berkeley, CA, 94720}
\ead{yfzhang@lbl.gov}

\begin{abstract}
We report on the measurement of D meson production from the
analysis of their hadronic ($D^0\rightarrow K\pi$) and
semileptonic ($D\rightarrow \mu+X$, $D\rightarrow e+X$) decays in
$\sqrt{s_{NN}}$=200 GeV Au+Au collisions. The transverse momentum
($p_T$) spectra and the nuclear modification factors for $D^0$ and
for electron/muon\footnote[2]{The word ``muon'' refers to
$\mu^+/\mu^-$, ``electron'' refers to electron/positron throughout
these proceedings.} from charm semileptonic decays will be
presented. The differential cross section $d\sigma/dy$ is found to
be consistent with the number of binary scaling. The blast-wave
fit suggests that the charm hadron freeze out earlier than other
light flavor hadrons.
\end{abstract}
\vspace{-0.5cm}

\section{Introduction}
\vspace{-0.35cm}

In relativistic heavy-ion collisions, charm quarks are predicted
to lose less energy compared to light quarks in the partonic
matter due to the ``dead-cone" effect~\cite{dead,miko,armesto}.
The $p_T$ distributions and nuclear modification factors of D
mesons and of single electrons from charmed hadron decay will be
vital to study the physics in relativistic heavy-ion
collisions~\cite{CharmSQM06}. Charm quarks are believed to be
produced at early stages via initial gluon fusion and their
production cross section can be evaluated by perturbative
QCD~\cite{cacciari}. Study of the binary collision ($N_{bin}$)
scaling properties for the charm total cross section among p+p,
d+Au to Au+Au collisions can test if heavy-flavor quarks are
produced exclusively at initial impact~\cite{ffcharm}. Due to the
heavy mass of charm quark, charmed hadrons might freeze out
earlier than light flavor hadrons. Their flow velocity might be a
good indicator of light flavor thermalization occurring at the
partonic level~\cite{therm,xu2,teaney,batsouli}.

\section{Analysis and Results}
\vspace{-0.35cm}

The data used for this analysis methods were taken with the STAR
experiment during the $\sqrt{s_{NN}}$=200 GeV Au+Au run in 2004. A
total of 13.3 and 7.8 million 0-80\% minimum bias Au+Au events
were used for the $D^0$ reconstruction and the Time-of-Flight
(TOF) single muon/electron analysis, respectively. About 15
million top 12\% central Au+Au collision events were also used for
TOF single muon and electron analysis.

The $D^0$ mesons ($p_{T}<3$ GeV/$c$) were reconstructed through
their decay $D^0\rightarrow K^-\pi^+$ ($\bar{D^0}\rightarrow
K^+\pi^-$) with a branching ratio of 3.83\%. The $D^0$ yields were
obtained from the invariant mass distributions of kaon-pion pairs
after mixed-event background subtraction. Analysis details can be
found in Ref.~\cite{dAuCharm}.
\begin{figure}
\centering
\includegraphics[height=3.1in,width=2.3in]{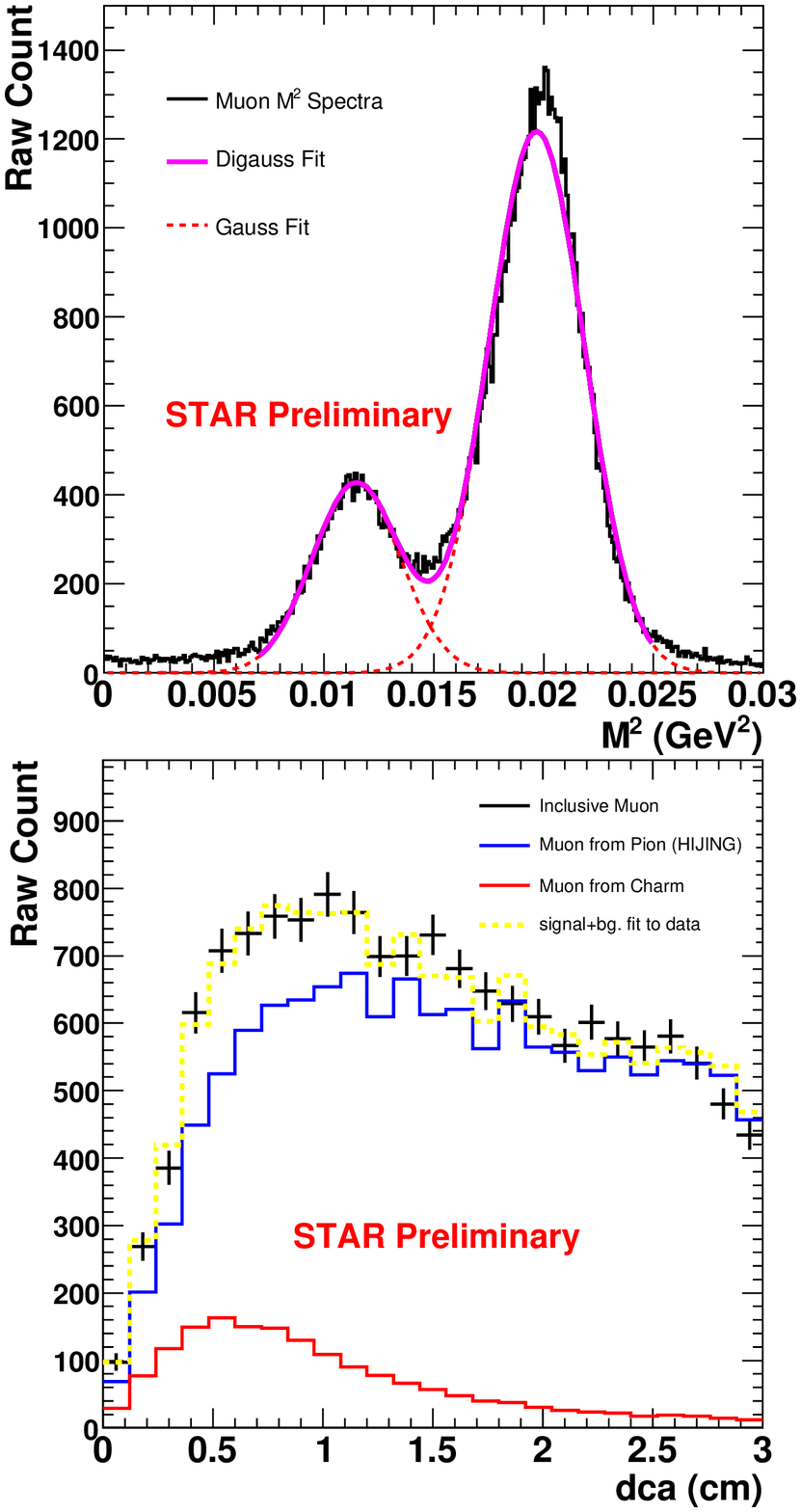}
\includegraphics[width=2.5in]{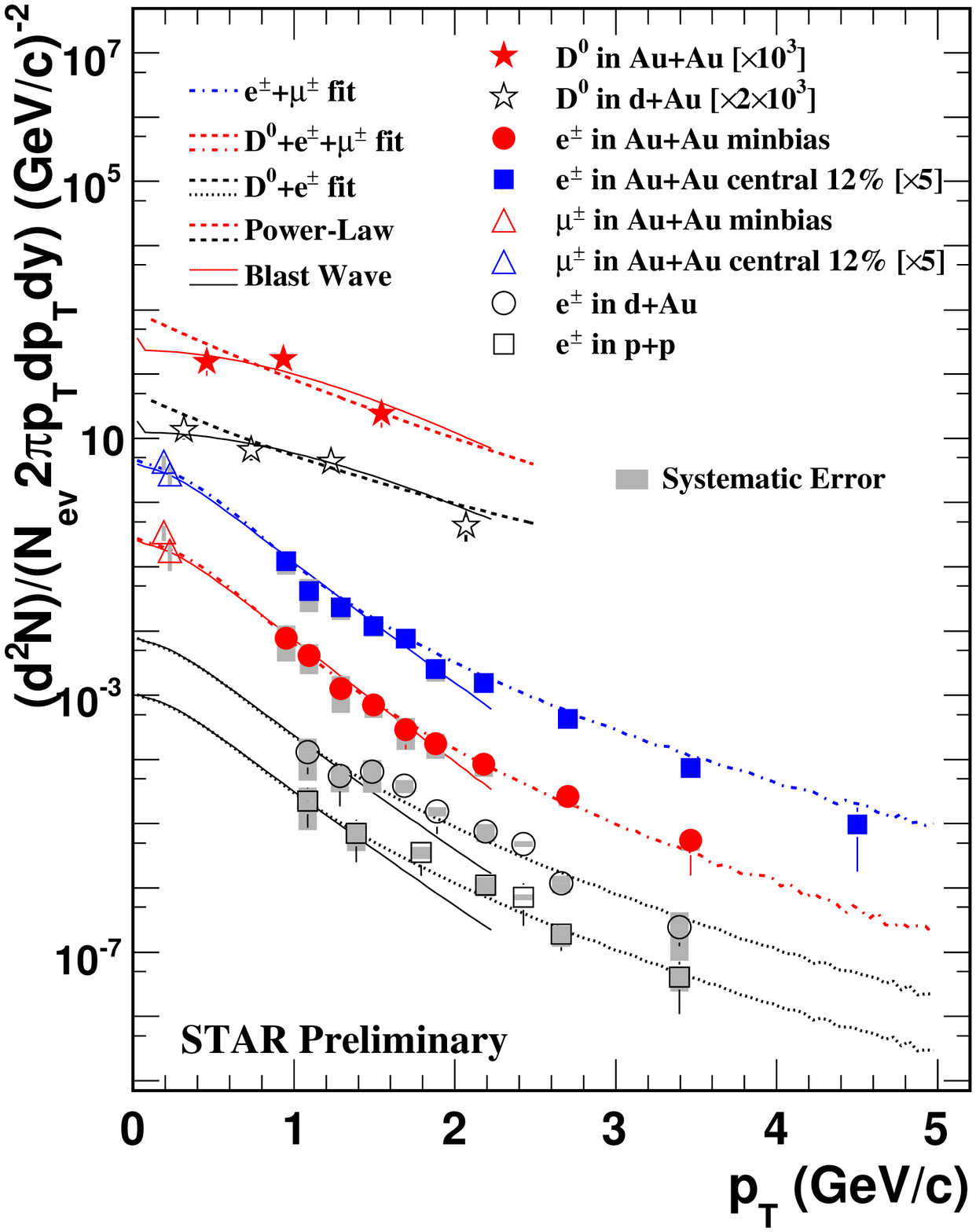}
\caption{Left panel: Upper part: Particle mass squared
distribution ($m^{2}=(p/\beta/\gamma)^{2}$) from the TOF after TPC
$dE/dx$ selections. A clear muon mass peak is observed and the
primary pion candidates are shown as the right peak. Bottom part:
Primary particle DCA (red line) and muon DCA distributions from
pion kaon weak decayed background after TPC $dE/dx$ and TOF
$m^{2}$ selections from HIJING simulation through realistic STAR
detector configuration (blue line). Right panel: The pT spectra of
$D^0$ and electron/muon from semileptonic decays in Au+Au and d+Au
collisions. Dashed curves are from the power-law fits by combining
above three measurements, while solid curves are from the
blast-wave fits by combining $D^0$ and muon spectra with
non-photonic electron spectra at $p_T<2$
GeV/$c$.}\vspace{-0.25cm}\label{fig:Figure1}
\end{figure}

Inclusive electrons can be identified up to $p_{T}=5$ GeV/$c$ by
using a combination of velocity ($\beta$) from the TOF and
ionization energy loss ($dE/dx$) measured in the STAR Time
Projection Chamber (TPC). To measure the photonic electron
spectra, the invariant mass and opening angle of the $e^+e^-$
pairs were constructed from an electron (positron) in the TOF at
lower $p_{T}$ ($<1.2$ GeV/$c$) or in the TPC at higher $p_{T}$
($1.2<p_{T}<5$ GeV/$c$) combined with every other positron
(electron) candidate reconstructed in the
TPC~\cite{johnson,xdong}. The non-photonic single electron spectra
($0.9<p_{T}<5$ GeV/$c$) from charm semileptonic decays can be
extracted from the inclusive electron spectra subtracted by
photonic background. Detailed analysis can be found in
Ref.~\cite{dAuCharm,xdong}.

\begin{figure}[htp]
\centering
\includegraphics[width=2.8in]{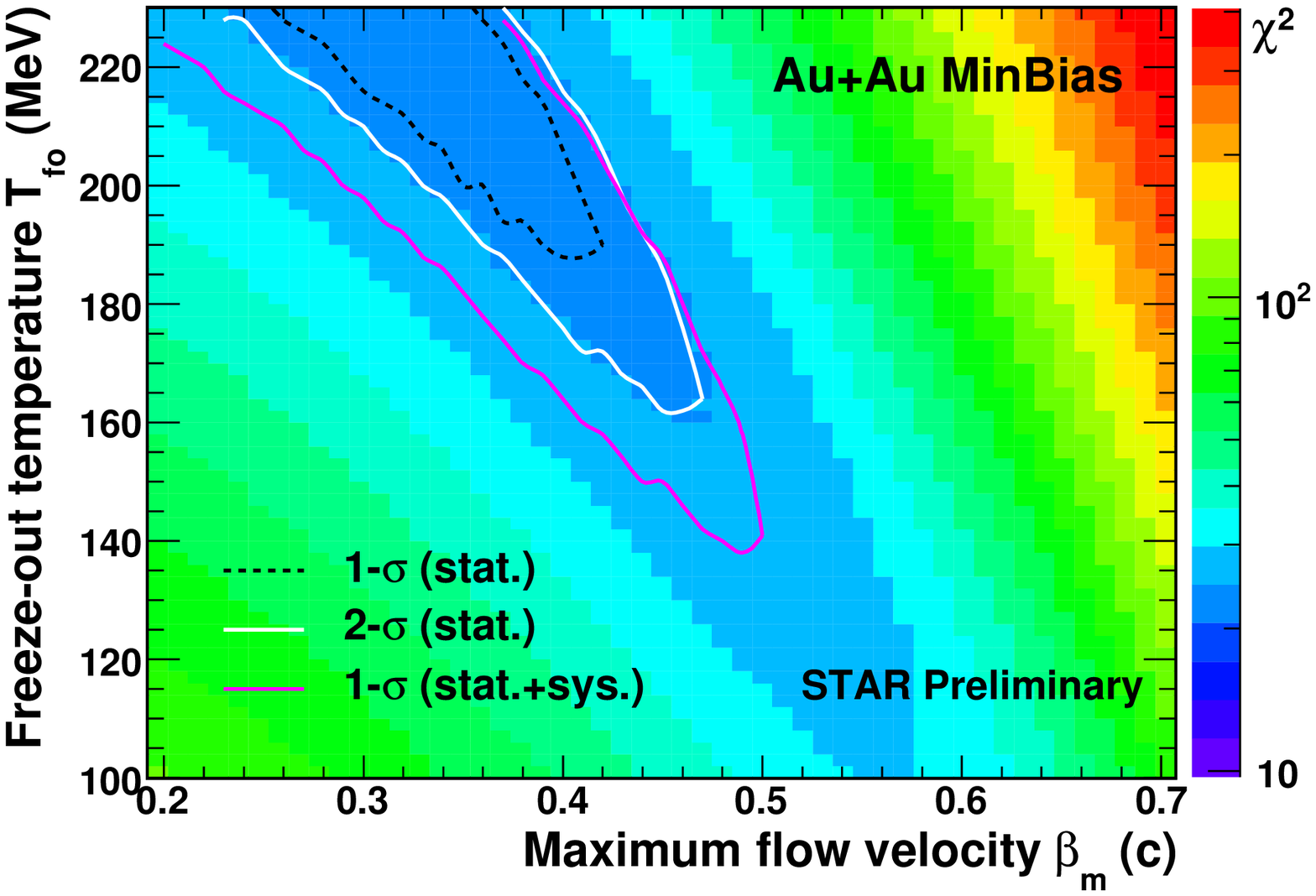}
\includegraphics[width=2.2in]{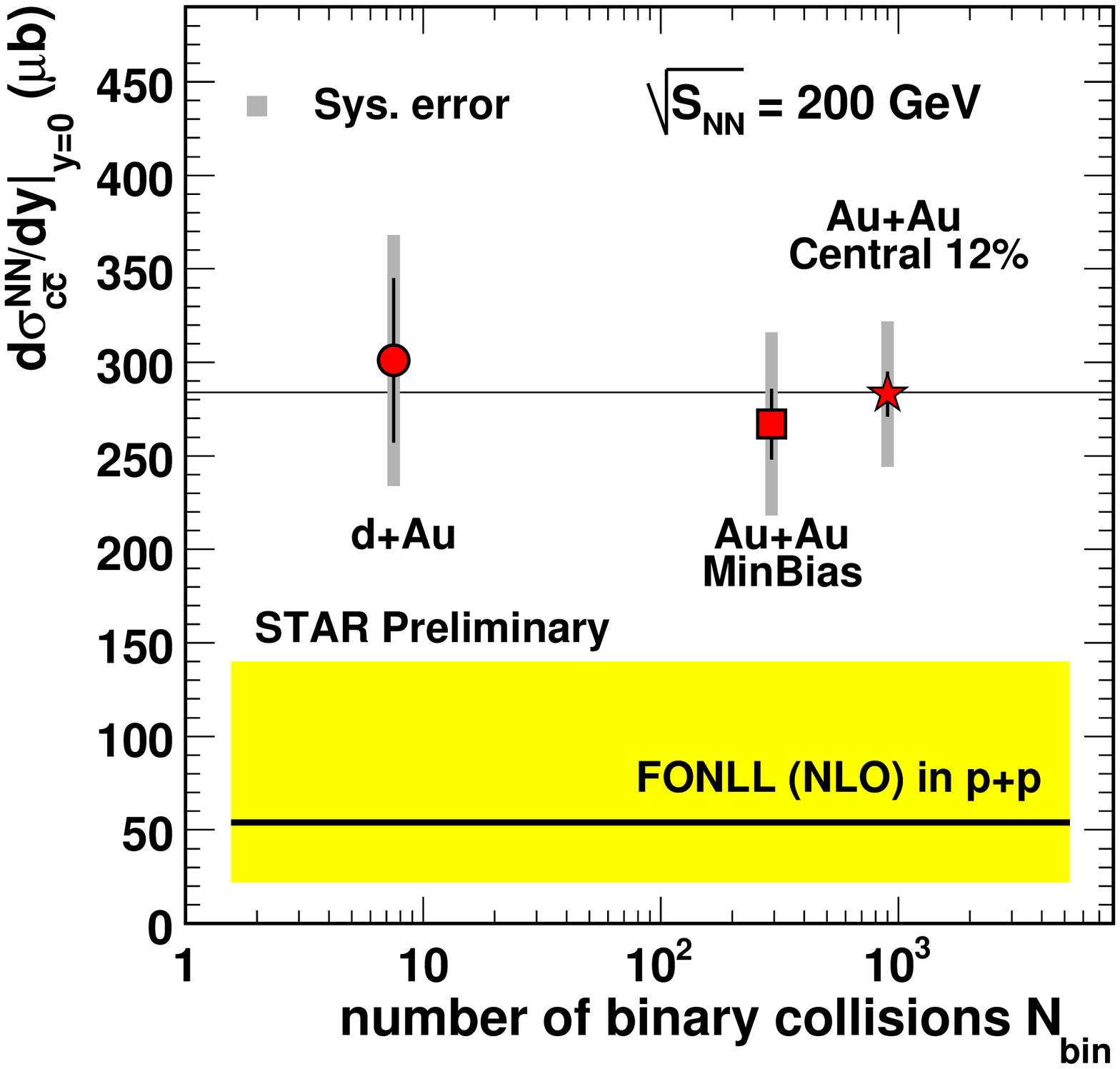}
\caption{Left panel: Charm hadron freeze-out temperature $T_{fo}$
versus maximum flow velocity $\beta_{m}$. Right panel: Charm cross
section at mid-rapidity compared with pQCD calculations in d+Au
and Au+Au collisions.}\vspace{-0.25cm}\label{fig:Figure2}
\end{figure}

However, due to the large combinatorial background in charmed
hadronic decay channels and the overwhelming photon conversions in
the detector material, we conducted the measurement of single muon
spectra at low $p_{T}$ ($0.17<p_{T}<0.25$ GeV/$c$) in both 0-80\%
minimum bias and top 12\% central Au+Au collisions at
$\sqrt{s_{_{NN}}} =200$ GeV. Muon identification was provided by
$dE/dx$ from the TPC and velocity from the TOF. Background muons
from pion/kaon weak decays were subtracted using the distribution
of the distance-of-closest-approach (DCA) to the collision vertex.
The single muon raw yield was obtained from the fit to muon DCA
distributions by combining the background DCA distributions and
the primary particle DCA distributions~\cite{ffcharm,czhongSQM06}.

The spectra of $D^0$ and muons/electrons from charmed hadron
decays are shown in the right panel of Fig.~\ref{fig:Figure1}. The
electron yield relies more on $D^0$ spectrum shape with a wide
range of power-law parameters $n$ and $\langle p_T \rangle$ at
fixed charm total yield. A factor of 8 variation of electron yield
integrated above $p_T$ of 1.0 GeV/$c$ was seen, while muon yield
only changes within $\pm15\%$~\cite{ffcharm}. We therefore
conducted a measurement of muons at low $p_T$ in order to
constrain the charm cross section. The $\langle p_T \rangle$ and n
can be derived from a power-law fit to the $D^0$ $p_T$
distributions and decayed lepton spectra shape, while the
freeze-out temperature $T_{fo}$ and flow velocity $\beta_{T}$ can
be derived from a blast-wave fit to the $D^0$ $p_T$ distributions
and decayed lepton spectra shape ($p_T<2$ GeV/$c$), see the right
panel of Fig.~\ref{fig:Figure1}. The $1\sigma$ contour from the
combined blast-wave fit with a quadratic sum of stat. and sys.
errors for the spectra shows the lower limit of charm hadron
$T_{fo}$ ($>140$ MeV) and small but non-zero $\beta_{T}$
($0.21\pm0.04(stat.)\pm0.07(sys.)$). This result suggests that
charm hadron seems to freeze out earlier than other hadrons, see
the left panel of Fig.~\ref{fig:Figure2}. The charm cross sections
at mid-rapidity ($d\sigma/dy$) can be obtained from the average of
the two fits by combining the three measurements covering
$\sim90\%$ of the kinematics. They are
$301\pm44(stat.)\pm67(sys.)\mu b$ in 200 GeV d+Au
collisions~\cite{dAuCharm}, $267\pm19\pm49\mu b$ in 200 GeV 0-80\%
minimum bias Au+Au collisions and $283\pm12\pm39\mu b$ in 200 GeV
top 12\% central Au+Au collisions. Within error bars, the measured
charm differential cross sections are found to be consistent with
the number of binary scaling, see the right panel of
Fig.~\ref{fig:Figure2}. In addition, the measured cross sections
are larger than the pQCD prediction by a factor of
5~\cite{cacciari}! Note that the systematic errors are the
dominant ones. In the fitting procedure, the statistical and
systematic errors were summed up quadratically.


\section{Conclusions}
\vspace{-0.35cm}

The measurement of charm production from analysis of
$D^0\rightarrow K\pi$, muon, and electron channels in both minimum
bias and central Au+Au collisions at RHIC was reported.

The transverse momentum spectra from non-photonic electrons are
strongly suppressed at $1<p_T<5$ GeV/$c$ in Au+Au collisions
relative to that in p+p and d+Au collisions~\cite{CharmSQM06}. For
electron with $p_T>2$ GeV/$c$, corresponding to charm hadron with
$p_T>4$ GeV/$c$, the suppression is similar to those of light
baryon and meson hadrons~\cite{lqeloss}.

Detailed model-dependent analysis of the electron spectra with
$p_T \le 2$ GeV/$c$ suggests that charm hadrons have a different
freeze-out pattern from the more copiously produced light hadrons.
The blast-wave fits show that charm hadrons seem to freeze out
early, and have similar freeze-out temperature as multistrangeness
hadrons ($\phi$, $\Omega$) but with smaller collective velocity.

Charm cross sections at mid-rapidity are extracted from a
combination of the three measurements covering $\sim90\%$ of the
total yield at mid-rapidity. The cross section is found to follow
binary scaling, which is a signature of charm production
exclusively at the initial impact. This supports the assumption
that hard processes scale with binary interactions among initial
nucleons and charm quarks can be used as a probe sensitive to the
early dynamical stage of the system.


\section*{References}
\vspace{-0.35cm}
\begin{thebibliography}{10}

\bibitem{dead} Y. Dokshizer et al. Phys. Lett. B 519 (2001) 199.

\bibitem{miko} M. Djordjevic et al. Phys. Rev. Lett. 94 (2005) 112301.

\bibitem{armesto} N. Armesto et al. Phys. Rev. D 71 (2005) 054027.

\bibitem{CharmSQM06}
H.B. Zhang {\it et al.} these proceedings.

\bibitem{therm}
STAR Collaboration, J. Adams {\it et al.},
\Journal{\PRL}{92}{112301}{2004}

\bibitem{xu2} N. Xu and Z. Xu \Journal{\NPA}{715}{587c}{2003};
    Z.W. Lin and D. Molnar, \Journal{\PRC}{68}{044901}{2003};
    V. Greco, C.M. Ko and R. Rapp, \Journal{\PLB}{595}{202}{2004}.

\bibitem{teaney} G.D. Moore, D. Teaney, \Journal{\PRC}{71}{064904}{2005}

\bibitem{batsouli}
S. Batsouli {\it et al.}, \Journal{PLB}{557}{2003}{26-32} e-print
Arxiv: nucl-th/0212068

\bibitem{cacciari} M. Cacciari, P. Nason and R. Vogt,
\Journal{\PRL}{95}{2005}{122001}

\bibitem{ffcharm} H.D. Liu et al. e-print Arxiv: nucl-ex/0601030

\bibitem{dAuCharm}
STAR Collaboration, J. Adams {\it et al.},
\Journal{\PRL}{94}{062301}{2005} e-print Arxiv: nucl-ex/0407006

\bibitem{johnson} J. Adams {\it et al.} (STAR Collaboration),
\Journal{\PRC}{70}{044902}{2004}; I.  Johnson, Ph.D. thesis, U.C.
Davis, 2002.

\bibitem{xdong} X. Dong, Ph.D. thesis, e-print Arxiv:
nucl-ex/0509011

\bibitem{czhongSQM06}
C. Zhong {\it et al.} these proceedings.

\bibitem{pdgcharmff} S. Eidelman {\it et al.}, \Journal{\PLB}{592}{1}{2004} (Particle Data Group);
R.M. Barnett {\it et al.}, \Journal{\PRD}{54}{486}{1996}.

\bibitem{lqeloss} STAR Collaboration, J. Adams {\it et al.}, e-print Arxiv: nucl-ex/0606003

\end{thebibliography}
\end{document}